\documentstyle [12pt] {article}

\catcode`\@=11
\def\@citex[#1]#2{%
\if@filesw \immediate \write \@auxout {\string \citation {#2}}\fi
\@tempcntb\m@ne \let\@h@ld\relax \def\@citea{}%
\@cite{%
  \@for \@citeb:=#2\do {%
    \@ifundefined {b@\@citeb}%
      {\@h@ld\@citea\@tempcntb\m@ne{\bf ?}%
      \@warning {Citation `\@citeb ' on page \thepage \space
undefined}}%
      {\@tempcnta\@tempcntb \advance\@tempcnta\@ne%
      \@tempcntb\number\csname b@\@citeb \endcsname \relax%
      \ifnum\@tempcnta=\@tempcntb 
	\ifx\@h@ld\relax%
	  \edef \@h@ld{\@citea\csname b@\@citeb\endcsname}%
	\else%
	  \edef\@h@ld{\ifmmode{-}\else--\fi\csname
b@\@citeb\endcsname}%
	\fi%
      \else
	\@h@ld\@citea\csname b@\@citeb \endcsname%
	\let\@h@ld\relax%
      \fi}%
    \def\@citea{,\penalty\@highpenalty\,}%
  }\@h@ld
}{#1}}

\def\@citeb#1#2{{[#1]\if@tempswa , #2\fi}}
\def\@citeu#1#2{{$^{#1}$\if@tempswa , #2\fi }}
\def\@citep#1#2{{#1\if@tempswa , #2\fi}}

\def\bcites{         
	\catcode`\@=11
	\let\@cite=\@citeb
	\catcode`\@=12
}

\def\upcites{         
	\catcode`\@=11
	\let\@cite=\@citeu
	\catcode`\@=12
}

\def\plaincites{      
	\catcode`\@=11
	\let\@cite=\@citep
	\catcode`\@=12
}

\newcount\hour
\newcount\minute
\newtoks\amorpm
\hour=\time\divide\hour by 60
\minute=\time{\multiply\hour by 60 \global\advance\minute by-\hour}
\edef\standardtime{{\ifnum\hour<12 \global\amorpm={am}%
	\else\global\amorpm={pm}\advance\hour by-12 \fi
	\ifnum\hour=0 \hour=12 \fi
	\number\hour:\ifnum\minute<10
0\fi\number\minute\the\amorpm}}
\edef\militarytime{\number\hour:\ifnum\minute<10
0\fi\number\minute}

\def\draftlabel#1{{\@bsphack\if@filesw {\let\thepage\relax
   \xdef\@gtempa{\write\@auxout{\string
      \newlabel{#1}{{\@currentlabel}{\thepage}}}}}\@gtempa
   \if@nobreak \ifvmode\nobreak\fi\fi\fi\@esphack}
	\gdef\@eqnlabel{#1}}
\def\@eqnlabel{}
\def\@vacuum{}
\def\marginnote#1{}
\def\draftmarginnote#1{\marginpar{\raggedright\scriptsize\tt#1}}
\def\draft{
	\pagestyle{plain}
	\overfullrule=2pt
	\oddsidemargin -.5truein
	\def\@oddhead{\sl \phantom{\today\quad\militarytime} \hfil
	\smash{\Large\sl DRAFT} \hfil \today\quad\militarytime}
	\let\@evenhead\@oddhead
	\let\label=\draftlabel
	\let\marginnote=\draftmarginnote
	\def\ps@empty{\let\@mkboth\@gobbletwo
	\def\@oddfoot{\hfil \smash{\Large\sl DRAFT} \hfil}
	\let\@evenfoot\@oddhead}

\def\@eqnnum{(\theequation)\rlap{\kern\marginparsep\tt\@eqnlabel}%
	\global\let\@eqnlabel\@vacuum}  }

\def\blackfonts{
	\font\blackboard=msbm10 scaled\magstep1
	\font\blackboards=msbm8
	\font\blackboardss=msbm6
}

\def\nblack{            
	\def\ZZ{{Z \n{10} Z}}
	\def\NN{{N \n{14} N}}
	\def\CC{{C \n{11} C}}
	\def\RR{{R \n{11} R}}
	\def\QQ{{Q \n{12} Q}}
	\def\PP{{P \n{11} P}}
}

\def\prep{         
	\catcode`\@=11
	\input art10.sty
	\catcode`\@=12
	
	\let\small\null
	\def\blackfonts{
		\font\blackboard=msbm10
		\font\blackboards=msbm7
		\font\blackboardss=msbm5
	}
	\let\sl\it
	\twocolumn
	\sloppy
	\voffset=-2.54truecm
	\hoffset=-2.54truecm
	\flushbottom
	\parindent 1em
	\leftmargini 2em
	\leftmarginv .5em
	\leftmarginvi .5em
	\marginparwidth 48pt
	\marginparsep 10pt
	\setlength{\columnsep}{2truecm}
	\setlength{\textwidth}{25.4truecm}
	\setlength{\textheight}{17truecm}
>	\baselineskip=16pt
	\oddsidemargin .18truein
	\evensidemargin .17truein
}

\def\eqalign#1{\null\,\vcenter{\openup\jot\m@th
  \ialign{\strut\hfil$\displaystyle{##}$&$\displaystyle{{}##}$\hfil
      \crcr#1\crcr}}\,}
\def\eqalignno#1{\displ@y \tabskip\centering
  \halign
to\displaywidth{\hfil$\@lign\displaystyle{##}$\tabskip\z@skip
    &$\@lign\displaystyle{{}##}$\hfil\tabskip\centering
    &\llap{$\@lign##$}\tabskip\z@skip\crcr
    #1\crcr}}

\def\section{\@startsection {section}{1}{\z@}{3.ex plus 1ex minus
 .2ex}{2.ex plus .2ex}{\large\bf}}
\def\subsection{\@startsection{subsection}{2}{\z@}{2.75ex plus 1ex
minus
 .2ex}{1.5ex plus .2ex}{\bf}}

\def\appendix{{\newpage\section*{Appendix}}\let\appendix\section%
	{\setcounter{section}{0}
	\gdef\thesection{\Alph{section}}}\section}

\def\abstract{\if@twocolumn
\section*{Abstract}
\else 
\begin{center}
{\bf Abstract\vspace{-.5em}\vspace{0pt}}
\end{center}
\quotation
\fi}

\catcode`\@=12

\def\ibid{{\it ibid.\/}}

\newcommand{\beq}{\begin{equation}}
\newcommand{\eeq}{\end{equation}}
\newcommand{\beqa}{\begin{eqnarray}}
\newcommand{\eeqa}{\end{eqnarray}}

\def\noj#1,#2,{{\bf #1} (19#2)\ }
\def\jou#1,#2,#3,{{\sl #1\/ }{\bf #2} (19#3)\ }
\def\ann#1,#2,{{\sl Ann.\ Physics\/ }{\bf #1} (19#2)\ }
\def\cmp#1,#2,{{\sl Comm.\ Math.\ Phys.\/ }{\bf #1} (19#2)\ }
\def\ma#1,#2,{{\sl Math.\ Ann.\/ }{\bf #1} (19#2)\ }
\def\jd#1,#2,{{\sl J.\ Diff.\ Geom.\/ }{\bf #1} (19#2)\ }
\def\invm#1,#2,{{\sl Invent.\ Math.\/ }{\bf #1} (19#2)\ }
\def\cq#1,#2,{{\sl Class.\ Quantum Grav.\/ }{\bf #1} (19#2)\ }
\def\cqg#1,#2,{{\sl Class.\ Quantum Grav.\/ }{\bf #1} (19#2)\ }
\def\ijmp#1,#2,{{\sl Int.\ J.\ Mod.\ Phys.\/ }{\bf A#1} (19#2)\ }
\def\jmphy#1,#2,{{\sl J.\ Geom.\ Phys.\/ }{\bf #1} (19#2)\ }
\def\jams#1,#2,{{\sl J.\ Amer.\ Math.\ Soc.\/ }{\bf #1} (19#2)\ }
\def\grg#1,#2,{{\sl Gen.\ Rel.\ Grav.\/ }{\bf #1} (19#2)\ }
\def\mpl#1,#2,{{\sl Mod.\ Phys.\ Lett.\/ }{\bf A#1} (19#2)\ }
\def\nc#1,#2,{{\sl Nuovo Cim.\/ }{\bf #1} (19#2)\ }
\def\np#1,#2,{{\sl Nucl.\ Phys.\/ }{\bf B#1} (19#2)\ }
\def\pl#1,#2,{{\sl Phys.\ Lett.\/ }{\bf #1B} (19#2)\ }
\def\pla#1,#2,{{\sl Phys.\ Lett.\/ }{\bf #1A} (19#2)\ }
\def\pr#1,#2,{{\sl Phys.\ Rev.\/ }{\bf #1} (19#2)\ }
\def\prd#1,#2,{{\sl Phys.\ Rev.\/ }{\bf D#1} (19#2)\ }
\def\prl#1,#2,{{\sl Phys.\ Rev.\ Lett.\/ }{\bf #1} (19#2)\ }
\def\prp#1,#2,{{\sl Phys.\ Rept.\/ }{\bf #1C} (19#2)\ }
\def\ptp#1,#2,{{\sl Prog.\ Theor.\ Phys.\/ }{\bf #1} (19#2)\ }
\def\ptpsup#1,#2,{{\sl Prog.\ Theor.\ Phys.\/ Suppl.\/ }{\bf #1}
(19#2)\ }
\def\rmp#1,#2,{{\sl Rev.\ Mod.\ Phys.\/ }{\bf #1} (19#2)\ }
\def\yadfiz#1,#2,#3[#4,#5]{{\sl Yad.\ Fiz.\/ }{\bf #1} (19#2) #3%
\ [{\sl Sov.\ J.\ Nucl.\ Phys.\/ }{\bf #4} (19#2) #5]}
\def\zh#1,#2,#3[#4,#5]{{\sl Zh..\ Exp.\ Theor.\ Fiz.\/ }{\bf #1}
(19#2) #3%
\ [{\sl Sov.\ Phys.\ JETP\/ }{\bf #4} (19#2) #5]}

\def\beq{\begin{equation}}
\def\eeq{\end{equation}}
\def\beqar{\begin{eqnarray}}
\def\eeqar{\end{eqnarray}}

\def\nfrac#1#2{{\displaystyle{\vphantom1\smash{\lower.5ex\hbox{\sma
ll$#1$}}%
	\over\vphantom1\smash{\raise.25ex\hbox{\small$#2$}}}}}

\def\n#1{\mskip-#1mu}

\def\lae{\mathrel{\mathop{\smash{\lower .5 ex
\hbox{$\stackrel<\sim$}}}}}
\def\lae{\mathrel{\mathop{\smash{\lower .5 ex
\hbox{$\stackrel>\sim$}}}}}

\def\l:{\mathopen{:}\,}
\def\r:{\,\mathclose{:}}

\catcode`\@=11
\def\theequation{\arabic{equation}}
\catcode`\@=12

\nblack
\bcites

\nblack

\catcode`\@=11
\def\theequation{\thesection.\arabic{equation}}
\@addtoreset{equation}{section}
\@addtoreset{footnote}{section}
\@addtoreset{footnote}{subsection}
\catcode`\@=12

\newcommand{\beqn}{\begin{equation}}
\newcommand{\eeqn}{\end{equation}}
\newcommand{\beqnarray}{\begin{eqnarray}}
\newcommand{\eeqnarray}{\end{eqnarray}}
\newcommand {\bear} [1] {\begin {array} {#1}}
\newcommand {\ear} {\end {array}}

\newcommand {\beqarn} {\begin{eqnarray*}}
\newcommand {\eeqarn} {\end{eqnarray*}}

\newcommand {\mrm} [1] {\mathrm {#1}}

\newcommand {\myref} [1]	%
	{%
	\begin{thebibliography} {99}	%
			{#1}	%
	\end {thebibliography}}

\def\emph#1{{\em #1}}
\def\mathbf#1{{\bf #1}}
\def\mathrm#1{{\rm #1}}
\def\mathit#1{{\it #1}}

\catcode`\@=11
\@addtoreset{footnote}{section}
\catcode`\@=12

\newcounter	{exinsert}	[subsection]

\renewcommand {\theexinsert}	%
{	\thesubsection.\arabic{equation}}

\renewcommand {\theequation} {\thesection.\arabic{equation}}

\catcode`\@=11
\@addtoreset{equation}{section}
\catcode`\@=12

\newenvironment {exinsert} [1]	%
{	
	\begin {quotation}	%
	\refstepcounter {equation}	%
	{\bf {#1}} \theequation. \,\,	%
}
{	
	\end {quotation}
}

\newcommand {\bprop}
{\begin {exinsert} {Proposition}
}
\newcommand {\eprop} {\end {exinsert} }

\newcommand {\bexe} {\begin {exinsert} {Exercise}}
\newcommand {\eexe} {\end {exinsert} }

\newcommand {\bexa} {\begin {exinsert} {Example}}
\newcommand {\eexa} {\end {exinsert} }

\newcommand {\literal} [1] {{ \mbox {$\mrm {#1}$}}}

\newcommand {\grad} {\nabla}

\newcommand {\ncmd} [2] {\newcommand {#1} {#2}}

\ncmd {\vgrad} {{\vec \grad}}

\ncmd {\Mv} {{\cal M}_V}
\ncmd {\Mh} {{\cal M}_H}

\ncmd {\adj} {\literal {adj}}	
\ncmd {\Mp} {M_\literal {planck}}	

\ncmd {\x} {$\times$}

\newcommand {\btab} [3]
  {\begin{table} [htb]
   \begin{center}
   \caption {#2}	\label {tab:#1}
   \begin {tabular} {#3}
   \hline}

\newcommand {\etab}
	{  \hline   \end {tabular}
   \end{center}
  \end{table}}

\ncmd {\bR} {\textbf {R}}
\ncmd {\bC} {\textbf {C}}

\ncmd {\CQ} {\cal Q}

\def\jou#1,#2,#3,{{\sl #1\/ }{\bf #2} (19#3)\ }
\def\ibid#1,#2,{{\sl ibid.\/ }{\bf #1} (19#2)\ }
\def\am#1,#2,{{\sl Acta. Math.\/ } {\bf #1} (19#2)\ }
\def\annp#1,#2,{{\sl Ann.\ Physics\/ }{\bf #1} (19#2)\ }
\def\cmp#1,#2,{{\sl Comm.\ Math.\ Phys.\/ }{\bf #1} (19#2)\ }
\def\cqg#1,#2,{{\sl Class.\ Quantum Grav.\/ }{\bf #1} (19#2)\ }
\def\dm#1,#2,{{\sl Duke\ Math.\ J.\/ }{\bf #1} (19#2)\ }
\def\ijmpa#1,#2,{{\sl Int.\ J.\ Mod.\ Phys.\/ }{\bf A#1} (19#2)\ }
\def\invm#1,#2,{{\sl Invent.\ Math.\/ }{\bf #1} (19#2)\ }
\def\jhep#1,#2,{{\sl JHEP\/ }{\bf #1} (19#2)\ }
\def\jmp#1,#2,{{\sl J.\ Geom.\ Phys.\/ }{\bf #1} (19#2)\ }
\def\jams#1,#2,{{\sl J.\ Diff.\ Geom.\/ }{\bf #1} (19#2)\ }
\def\jdg#1,#2,{{\sl J.\ Amer.\ Math.\ Soc.\/ }{\bf #1} (19#2)\ }
\def\jpa#1,#2,{{\sl J.\ Phys.\ A.\/ }{\bf #1} (19#2)\ }
\def\grg#1,#2,{{\sl Gen.\ Rel.\ Grav.\/ }{\bf #1} (19#2)\ }
\def\mpla#1,#2,{{\sl Mod.\ Phys.\ Lett.\/ }{\bf A#1} (19#2)\ }
\def\ma#1,#2,{{\sl Math.\ Ann.\/ } {\bf #1} (19#2)\ }
\def\nc#1,#2,{{\sl Nuovo\ Cim.\/ }{\bf #1} (19#2)\ }
\def\npb#1,#2,{{\sl Nucl.\ Phys.\/ }{\bf B#1} (19#2)\ }
\def\plb#1,#2,{{\sl Phys.\ Lett.\/ }{\bf #1B} (19#2)\ }
\def\pla#1,#2,{{\sl Phys.\ Lett.\/ }{\bf #1A} (19#2)\ }
\def\pnas#1,#2,{{\sl Proc..Nat.Acad.Sci.\/ }{\bf #1} (19#2)\ }
\def\prev#1,#2,{{\sl Phys.\ Rev.\/ }{\bf #1} (19#2)\ }
\def\prd#1,#2,{{\sl Phys.\ Rev.\/ }{\bf D#1} (19#2)\ }
\def\prl#1,#2,{{\sl Phys.\ Rev.\ Lett.\/ }{\bf #1} (19#2)\ }
\def\prpt#1,#2,{{\sl Phys.\ Rept.\/ }{\bf #1C} (19#2)\ }
\def\ptp#1,#2,{{\sl Prog.\ Theor.\ Phys.\/ }{\bf #1} (19#2)\ }
\def\ptpsup#1,#2,{{\sl Prog.\ Theor.\ Phys.\/ Suppl.\/ }{\bf #1} (19#2)\ }
\def\rmp#1,#2,{{\sl Rev.\ Mod.\ Phys.\/ }{\bf #1} (19#2)\ }
\def\sm#1,#2,{{\sl Selec. Math.\/ }{\bf #1} (19#2)\ }
\def\yadfiz#1,#2,#3[#4,#5]{{\sl Yad.\ Fiz.\/ }{\bf #1} (19#2) #3%
\ [{\sl Sov.\ J.\ Nucl.\ Phys.\/ }{\bf #4} (19#2) #5]}
\def\zh#1,#2,#3[#4,#5]{{\sl Zh.\ Exp.\ Theor.\ Fiz.\/ }{\bf #1} (19#2) #3%
\ [{\sl Sov.\ Phys.\ JETP\/ }{\bf #4} (19#2) #5]}

\begin {document}


\begin{titlepage}

\begin{center}
\hfill
UCB-PTH-98/35,~~
LBNL-41948,~~
NSF-ITP-98-070\\
\hfill hep-th/9806171

\vskip 1.5 cm
{\Large \bf Glueballs and Their Kaluza-Klein Cousins}
\vskip 1 cm
{\large Hirosi Ooguri, Harlan Robins and Jonathan Tannenhauser}\\
\vskip 1cm
{Department of Physics,
University of California at Berkeley,\\
Berkeley, CA 94720}

{Theoretical Physics Group, Mail Stop 50A-5101,\\
 Lawrence Berkeley National Laboratory, \\
Berkeley, CA 94720}

{Institute for Theoretical Physics,
University of California,\\
Santa Barbara, CA 93106}

\end{center}

\vskip 0.5 cm
\begin{abstract}
Spectra of glueball masses in non-supersymmetric Yang-Mills theory 
in three and four dimensions have recently been computed using 
the conjectured duality between superstring theory and large
$N$ gauge theory. The Kaluza-Klein states of supergravity
do not correspond to any states in the Yang-Mills theory and
therefore should decouple in the continuum limit. On the 
other hand, in the supergravity limit $g_{YM}^2 N \rightarrow \infty$, 
we find that the masses of the Kaluza-Klein states
are comparable to those of the glueballs. 
We also show that the leading $(g_{YM}^2N)^{-1}$ corrections 
do not make these states heavier than the glueballs. 
Therefore, the decoupling of the Kaluza-Klein states
is not evident to this order. 

\end{abstract}
\end{titlepage}

\section{Introduction}

Spectra of glueball masses in non-supersymmetric Yang-Mills theory in three
and four dimensions have recently been calculated \cite{coot} using
the conjectured duality between string theory and
large $N$ gauge theory \cite{mal,GKP,w1,w2}.  The results 
are apparently in good numerical agreement with available 
lattice gauge theory data, although a direct comparison may 
be somewhat subtle, since the supergravity computation 
is expected to be valid for large ultraviolet coupling
$\lambda = g_{YM}^2 N$, whereas we expect that QCD in the continuum
limit is realized for $\lambda \rightarrow 0$ \cite{w2,go}.  
As explained in \cite{go,coot}, the supergravity computation
at $\lambda \gg 1$ gives the glueball masses in units of
the fixed ultraviolet cutoff $\Lambda_{UV}$. For finite $\lambda$,
the glueball mass $M$ is expected to be a function of the form
\beq
M^2 = F(\lambda) \Lambda_{UV}^2 .
\eeq
In the continuum limit $\Lambda_{UV} \rightarrow \infty$, 
$M$ should remain finite and of order $\Lambda_{QCD}$. 
This would require $F(\lambda) \rightarrow 0$
as $\lambda \rightarrow 0$. 
In \cite{coot}, the leading string theory corrections 
to the masses were computed and shown
to be negative and of order $\lambda^{-3/2}$,
in accordance with expectation.

 Witten has proposed \cite{w2} that
three-dimensional pure QCD is dual to type IIB string
theory on the product of an AdS$_5$ black hole and ${\bf S}^5$.
This proposal requires 
that certain states in string theory decouple in the continuum limit
$\lambda \rightarrow 0$. One class of such states are
Kaluza-Klein excitations on ${\bf S}^5$. 
The supergravity fields on the AdS$_5$ black hole $\times ~ {\bf S}^5$ 
can be classified
by decomposing them into spherical harmonics (the Kaluza-Klein
modes) on ${\bf S}^5$ \cite{van0,gm}. They fall into irreducible 
representations of the isometry group $SO(6)$ of ${\bf S}^5$,
which is the $R$-symmetry of the four-dimensional ${\cal N}=4$
supersymmetric gauge theory from which QCD$_3$ is obtained
by compactification on a circle. 
Consequently, only $SO(6)$ singlet states should correspond to
physical states in QCD$_3$ in the continuum limit.
These are the glueball states studied in \cite{coot}. 
However, we find that, in the supergravity
limit, masses of the $SO(6)$ non-singlet states
are of the same order as
the $SO(6)$ singlet states. Since these states
should decouple in the limit $\lambda \rightarrow 0$,
it was speculated in \cite{coot} that the string theory
corrections should make the non-singlet states heavier than the
singlet states. 

The purpose of this paper is to test this idea.
We compute the masses of the $SO(6)$ non-singlet
states coming from the Kaluza-Klein excitations of the dilaton
in ten dimensions. We find
the masses in the supergravity limit to be
of the same order as those of the $SO(6)$ singlet states. 
We then calculate the leading string theory
corrections to the masses. We find that the leading
corrections do not
make the Kaluza-Klein states heavier than
the glueballs. Therefore, the decoupling of the Kaluza-Klein states
is not evident to this order. This suggests that
the quantitative agreement between the glueball masses
from supergravity and the lattice gauge theory data
should be taken with a grain of salt.

\section{The Supergravity Limit}

We calculate the masses of the Kaluza-Klein states
following the analysis of \cite{coot}. According to 
\cite{w2},  QCD$_3$ is dual to type IIB superstring theory
on the AdS$_5$ black hole $\times ~ {\bf S}^5$ geometry given by
\beq
  \frac{dx^2}{l_s^2 \sqrt{4 \pi g_{YM}^2 N}}
  = \frac{d\rho^2}{\left( \rho^2 - \frac{b^4}{\rho^2} \right)}
 + \left( \rho^2 - \frac{b^4}{\rho^2} \right) d\tau^2 
  + \rho^2 \sum_{i=1}^3 dx_i^3 + d \Omega_5^2,
\label{oldbh}
\eeq
where $d \Omega_5$ is the line element on the unit ${\bf S}^5$ and
$l_s$ is the string length. The horizon of the black hole is
located at $\rho = b$. In order for the geometry to
be regular at the horizon, the coordinate $\tau$ must be
periodic with period $2\pi R$, where
$R = (2b)^{-1}$. The inverse radius $R^{-1}$ serves as the ultraviolet
cutoff of QCD$_3$; namely, $\Lambda_{UV} = (2R)^{-1}= b $.

To compute the mass of an $SO(6)$ non-singlet state, 
we express the dilaton field $\Phi$ as
\beq
\Phi = f_0(\rho) e^{ikx} Y_l(\Omega_5),
\eeq
where $Y_l(\Omega_5)$ is the $l$-th spherical harmonic
on ${\bf S}^5$, and solve the dilaton equation
in the geometry (\ref{oldbh}). This equation reduces to a
second-order ordinary differential equation for $f_0(\rho)$;
in units in which $b=1$,
\beq
 \rho^{-1} \frac{d}{d\rho}
\left( (\rho^4 - 1) \rho \frac{df_0}{d\rho}
\right) - (k_0^2 + l(l+4) \rho^2)f_0 = 0.
\label{olddiff}
\eeq
The mass in three dimensions is equal to $-k_0^2$ \cite{w2}.
Since the geometry (\ref{oldbh}) is smooth everywhere,
we require that $f_0(\rho)$ be regular everywhere, and in particular
at $\rho = \infty$ and at the horizon $\rho = 1$. 
The equation admits a regular solution
$f_0(\rho)$ for discrete values of $k_0^2$. 
This determines the mass spectrum. 

As in \cite{coot}, we determine $M^2 = - k_0^2$ numerically
by the shooting method. We first solve the differential equation 
(\ref{olddiff}) as an asymptotic
expansion in $\rho^{-2}$ and compute the first few terms
in the expansion. We then numerically integrate the equation,
with boundary conditions derived from the asymptotic expansion
imposed at a sufficiently large value of $\rho$ ($\rho \gg k_0^2$). 
The solution must be regular at the
horizon $\rho = 1$. This requirement determines the spectrum of $k_0^2$.
In the numerical evaluation, we find it convenient
to set the boundary condition
to be $f_0' = 0$ at the horizon. As we will show in the Appendix, this
shooting method can be used to compute $k_0^2$ and the
wavefunction $f_0(\rho)$ to arbitrarily high precision. 
The results of the numerical work are listed in Table 1.
As expected, the masses are all of the order of the ultraviolet
cutoff $\Lambda_{UV} = b$.

\begin{table}[htbp]
\centering
\begin{tabular}{l|l|l|l|l|l|l|l|l}
$l$&0 & 1 & 2 & 3 & 4 & 5&6&7\\ \hline
$M_l^2$ & 11.59 & 19.43&29.26 & 41.10&54.93 &70.76&88.60 &108.4 \\
$M_l^{*2}$ & 34.53 & 48.07 & 63.60& 81.11& 100.6& 122.1&145.6&171.1\\
$M_l^{**2}$& 68.98 & 88.24&109.5 & 132.7& 157.9&185.1&214.3&245.5 \\
\end{tabular}
\label{summary1}
\parbox{4in}{\caption{3d (Mass)$^2$ of the $l$-th Kaluza-Klein
modes on ${\bf S}^5$ and their excited states, in units of $b^2$}}
\end{table}

\section{Leading String Theory Corrections}

Witten's proposal requires that the Kaluza-Klein
states decouple in the continuum limit
$\lambda \rightarrow 0$. Here we examine whether
this efffect is evident from the leading string theory corrections. 

According to \cite{GKT}, the leading $\alpha' = (4 \pi g_{YM}^2 N)^{-1/2}$
correction to the AdS$_5$ black hole metric is
\beq
 \frac{ds^2}{l_s^2\sqrt{4 \pi g_{YM}^2 N}}
 = (1 + \delta_1) \frac{d\rho^2 }{\left( \rho^2
 - \frac{b^4}{\rho^2} \right)}
 + (1 + \delta_2)\left( \rho^2 -
\frac{b^4}{\rho^2} \right) d \tau^2
 + \rho^2 \sum_{i=1}^3 dx_i^2 + d \Omega_5^2,
\label{metric}
\eeq
where 
\beqa
  \delta_1 &=& + 15 \gamma \left( 5 
\frac{b^4}{\rho^4} + 5 \frac{b^8}{\rho^8} 
 - 19 \frac{b^{12}}{\rho^{12}} \right) \nonumber\\
\delta_2 &=& - 15 \gamma \left( 5 
\frac{b^4}{\rho^4} + 5 \frac{b^8}{\rho^8} 
 - 3 \frac{b^{12}}{\rho^{12}} \right),
\eeqa
and $\gamma = \frac{1}{8}\zeta(3) \alpha'^3$. 
In this geometry, the dilaton is no longer constant, 
but is given by
\beq
  \Phi_0 = - \frac{45}{8} \gamma
\left( \frac{b^4}{\rho^4} 
 + \frac{b^8}{2 \rho^8}
  + \frac{b^{12}}{3\rho^{12}} \right).
\label{dilaton}
\eeq
There is also a correction to the ten-dimensional dilaton action
\cite{grisaru,gw},
\beq
  I_{dilaton}
 = - \frac{1}{16\pi G_{10}}\int d^{10} x
\sqrt{g} \left[ - \frac{1}{2} g^{\mu\nu} \partial_\mu \Phi
  \partial_\nu \Phi + \gamma e^{-\frac{3}{2} \Phi} W \right],
\label{action}
\eeq
where $W$ is given in terms of the Weyl tensor.
In our background and in units where $b=1$,
$W = 180/\rho^{16}$. The relation between the location of 
the horizon $\rho = b$ and the periodicity $2 \pi R$ 
of $\tau$ is also modified to
\beq 
   R = \left( 1 - \frac{15}{8} \zeta(3) \alpha'^3
  + \cdots \right)\frac{1}{2b}.
\label{cutoff}
\eeq
It is the inverse radius $R^{-1}$ that serves as the ultraviolet cutoff of
QCD$_3$.

To solve the dilaton wave equation in the $\alpha'$-corrected 
geometry (\ref{metric}), we write
\beq
\Phi = \Phi_0 + f(\rho) e^{ikx} Y_l(\Omega_5),
\eeq
where $\Phi_0$ is the dilaton background
given by (\ref{dilaton}), and expand $f(\rho)$
and $k^2$ in $\gamma$ as
\beq
  f(\rho) = f_0(\rho) + \gamma h(\rho),
~~ k^2 = k_0^2 + \gamma \delta k^2.
\eeq
Here $f_0(\rho)$ obeys the lowest order equation
(\ref{olddiff}) and is a numerically given function,
and $k_0^2$ is likewise determined from (\ref{olddiff}).
The second-order differential equation obtained
from the action (\ref{action}) in the background
metric (\ref{metric}) and dilaton field
(\ref{dilaton}) is, in units in which $b=1$, 
\begin{eqnarray}
&& \rho^{-1} {d \over d\rho} \left(
(\rho^4 - 1) \rho {d h \over d\rho} \right)
- (k_0^2 + l (l+4) \rho^2) h = \nonumber \\
&&= (75 - 240 \rho^{-8}  + 165 \rho^{-12})
{d^2 f_0 \over d\rho^2}  \nonumber \\
&& + (75 + 1680 \rho^{-8} - 1815 \rho^{-12}) \rho^{-1}
 {d f_0 \over d\rho} \nonumber \\
& &+ ( \delta k^2   - 120(  k^2_0 + l(l+4) \rho^2) 
\rho^{-12}  - 405 \rho^{-14}  ) f_0(\rho).
\label{inhomo}
\end{eqnarray}
With $f_0(\rho)$ and $k_0^2$ given, one may regard
this as an inhomogeneous version of the
equation (\ref{olddiff}). We solve this equation
for $h(\rho)$ and $\delta k^2$.

We are now ready to present our results. Let us denote
the lowest mass of the $l$-th Kaluza-Klein state by $M_l$.
In units of the ultraviolet cutoff $\Lambda_{UV}= (2R)^{-1}$,
with $R$ given by (\ref{cutoff}), we find 
\beqa
 M^2_0~ & = &  11.59\times (1 - 2.78 \zeta(3) \alpha'^3 + \cdots
) \Lambda_{UV}^2 \nonumber \\
 M^2_1~ & = &  19.43\times (1 - 2.66\zeta(3) \alpha'^3 + \cdots
) \Lambda_{UV}^2 \nonumber \\
 M^2_2~ & = &  29.26\times (1 -2.62 \zeta(3) \alpha'^3 + \cdots
) \Lambda_{UV}^2 \nonumber \\
 M^2_3~ & = &  41.10\times (1 -2.61 \zeta(3) \alpha'^3 + \cdots
) \Lambda_{UV}^2 \nonumber \\
 M^2_4~ & = &  54.93\times (1 -2.63 \zeta(3) \alpha'^3 + \cdots
) \Lambda_{UV}^2 \nonumber \\
 M^2_5~ & = &  70.76\times (1 -2.66\zeta(3) \alpha'^3 + \cdots
) \Lambda_{UV}^2 \nonumber \\
 M^2_6~ & = &  88.60\times (1 -2.69 \zeta(3) \alpha'^3 + \cdots
) \Lambda_{UV}^2 \nonumber \\
 M^2_7~ & = &  108.4\times (1 -2.72 \zeta(3) \alpha'^3 + \cdots
) \Lambda_{UV}^2. 
\eeqa
Similar behavior is observed for the excited levels of
each Kaluza-Klein state. 

Thus the corrections do not
make the Kaluza-Klein states heavier than
the glueballs, and the decoupling of the Kaluza-Klein states
is not evident to this order.
According to Maldacena's duality, 
the $\lambda^{-1/2}$ expansion of the gauge theory corresponds to the 
$\alpha'$-expansion of the two-dimensional sigma model
with the AdS$_5$ black hole $\times ~ {\bf S}^5$ as its
target space. It is possible that the decoupling of
the Kaluza-Klein states takes place only non-perturbatively
in the sigma model.

\section*{Acknowledgments}

We thank Csaba Cs\'aki, Aki Hashimoto,
Yaron Oz, John Terning, and especially  David Gross 
for useful discussions. 
We thank the Institute for Theoretical Physics
at  Santa Barbara for
its hospitality. 

This work was supported in part by the NSF
grant PHY-95-14797 and the DOE grant DE-AC03-76SF00098,
and in part by 
the NSF grant PHY-94-07194 through ITP. 
H.R. and J.T. gratefully acknowledge the support of the A. Carl Helmholz 
Fellowship in the Department of Physics at the University of California, 
Berkeley.

\section*{Appendix: The  Boundary Condition at the Horizon}

In this appendix, we show that the boundary condition
at the horizon $\rho = b$ used in the shooting method
\cite{coot} is consistent, and that
the eigenvalue $k^2$ and the wavefunction $f(\rho)$
can be evaluated 
to an arbitrarily high precision using this method. 

In the neighborhood of $\rho = b$, the dilaton wave equation
takes the form
\beq
    \partial_\rho (\rho - b) \partial_\rho f(\rho) + \cdots
 = 0. 
\eeq
Its general solution is of the form
\beq
  f(\rho) = c_1 \left[ 1 + \alpha (\rho- b)  + \cdots
\right] + c_2 \left[ {\rm log}(\rho-b) + 
  \cdots \right]
\label{nearhorizon}
\eeq
with arbitrary coefficients $c_{1,2}$ (the constant
$\alpha$ is determined by the wave equation and is in general
non-zero). The regularity of 
the dilaton field requires $c_2=0$. In the shooting method,
we numerically integrate the differential equation starting
from a sufficiently
large value of $\rho$ down to the horizon. For generic $k^2$,
the function thus obtained, when expanded as in (\ref{nearhorizon}),
would have $c_2 \neq 0$. The task 
is to adjust $k^2$ so that $c_2 = 0$.

Since $f(\rho)$ is divergent at $\rho=b$ for generic $k^2$,
it is numerically
difficult to impose the boundary condition directly at $\rho = b$. 
Instead, in \cite{coot} and in this
paper, we required $f'= 0$ at $\rho = b + \epsilon$ for 
a given small $\epsilon$ (for example, $\epsilon = 0.0000001 b$ 
in this paper). By (\ref{nearhorizon}), this condition
implies
\beq
     c_2 = - c_1 \alpha \epsilon + \cdots.
\eeq
Therefore, $c_2$ can be made arbitrarily small by
adjusting $\epsilon$. This justifies the numerical
method used in \cite{coot} and in this paper.

We thank Aki Hashimoto for discussions on the numerical method.

\end{document}